\documentclass[prl,a4paper,twocolumn,footinbib,showpacs]{revtex4}

\usepackage[latin1]{inputenc}
\usepackage{amssymb}
\usepackage{amsmath}
\usepackage{amsbsy}
\usepackage{bm}
\usepackage{graphicx}

\newcommand{\lz}{l_0}
\newcommand{\tz}{t_0}
\newcommand{\hz}{h_0}
\newcommand{\wu}{\omega_1}
\newcommand{\wdd}{\omega_2}
\newcommand{\wde}{\omega_2^*}

\newcommand{\tfmu}{\mathcal{F}^{-1}}
\newcommand{\vk}{\mathbf{k}}

\newcommand{\vecr}{\mathbf{r}}
\newcommand{\vecu}{\mathbf{u}}
\newcommand{\nf}{\nu_f}
\newcommand{\ns}{\nu_s}
\newcommand{\Ef}{E_f}
\newcommand{\Es}{E_s}

\newcommand{\cE}{\mathcal{E}}
\newcommand{\cEz}{\mathcal{E}^0}
\newcommand{\pe}{\! = \!}
\newcommand{\psim}{\! \sim \!}
\newcommand{\ppp}{\! < \!}
\newcommand{\ppg}{\! > \!}

\newcommand{\tb}{t}
\newcommand{\db}{\delta}
\newcommand{\cw}{c_w}
\newcommand{\calH}{\mathcal{H}}
\newcommand{\cF}{\mathcal{F}}

\newcommand{\fud}{\frac{1}{2}}

\newcommand{\hwl}{h^{wl}}
\newcommand{\hmax}{h^{max}}
\newcommand{\Fel}{\mathcal{F}^{\mbox{\scriptsize{{\rm{el}}}}}}
\newcommand{\Fsurf}{\mathcal{F}^{\mbox{\scriptsize{{\rm{surf}}}}}}
\newcommand{\Vf}{V^{f}}

\begin{document}

\title{Nonlinear evolution of a morphological instability
 in a strained epitaxial film}
\author{Jean-No\"{e}l Aqua}\email{jnaqua@irphe.univ-mrs.fr}
\altaffiliation{\'Ecole Centrale Marseille }
\author{Thomas Frisch}
\author{Alberto Verga}
\affiliation{Institut de Recherche sur les Phénomènes Hors
\'Equilibre, UMR 6594, Aix-Marseille Université, France}

\date{\today}

\begin{abstract}
A strained epitaxial film deposited on a deformable substrate
undergoes a morphological instability relaxing the elastic energy by
surface diffusion. The nonlinear and nonlocal dynamical equations of
such films with wetting interactions are derived and solved
numerically in two and three dimensions. Above some critical
thickness, the surface evolves towards an array of islands separated
by a wetting layer. The island chemical potential decreases with
its volume, so that the system experiences  a non-interrupted
coarsening described by power laws with a marked dimension
dependence.
\end{abstract}

 \maketitle

The dynamics of semiconductor thin films is under active scrutiny
due to its importance for both fundamental science and technological
applications \cite{PimpVill98,FreuSure04}. Indeed, thin film elastic
instabilities lead to the self-organization of nanostructures
\cite{revues1} potentially useful e.g. for
quantum dots, wires and electronic devices with specific confinement
properties \cite{revues2}. A notorious experimental example is Si/Ge
films on a Si substrate which exhibit a variety of
structures such as pre-pyramids, pyramids, domes and huts
\cite{exp,Flor00}. Such epitaxial films experience an elastic stress 
due to the misfit with the substrate which is relaxed by a morphological 
instability similar to the Asaro-Tiller-Grinfeld
thermodynamical instability in solid-liquid interfaces \cite{ATG}.
This instability was first observed in experiments in helium at low 
temperature \cite{ToriBali92} and more generally in various solid interfaces  \cite{JessPenn93,Poli00,Mull04}.

Although the evolution of epitaxial films involves many complex
phenomena regarding surface energy, intermixing and kinetic processes,
we focus here on the main effects ruling the dynamics of the
morphological instability in strained films. The dynamics is ruled here
by surface diffusion driven by the interplay between isotropic surface energy 
and elastic energy \cite{Srol89,SpenVoor91}. When the
film is infinitely thick or when the substrate is infinitely rigid,
different theoretical \cite{ChiuGao93,SpenDavi93} and numerical
\cite{YangSrol93,SpenMeir94,KassMisb94,XianE02} approaches revealed
finite-time singularities enforced by elastic stress concentration
which account for experiments in thick films
\cite{ToriBali92,JessPenn93} where dislocations can finally develop.
However, these models can not describe experiments of thin films in
the Stranski-Krastanov type of growth \cite{exp,Flor00} where the
surface organizes smoothly into islands separated by a wetting layer
and evolving with a coarsening dynamics under annealing \cite{Flor00}. A
crucial issue for these systems is the wetting  of the substrate by
the film \cite{ChiuGao95,Spen99} which is a good candidate for
regularizing the dynamics of the instability. Indeed, crack
singularities were circumvented near the instability threshold by
considering slope dependent wetting effects \cite{GoloDavi03}.
However, the interplay between elastic relaxation, surface energy
and wetting interactions is still under active study
\cite{TekaSpen04,PangHuan06} and the description of the long term
dynamics of the morphological instability in a thin strained
film is an open issue. In this Letter, we present a model based on
continuum elasticity which we solve numerically revealing the
existence of a non-interrupted island coarsening.


We consider specifically a three dimensional (3D) dislocation free film
deposited on a substrate with slightly different lattice parameters
and with a priori different isotropic elastic properties. During
annealing, the film shape $h(x,y,t)$ changes by surface diffusion
(no external flux nor evaporation). The boundary at $z \pe h(x,y,t)$
is free while the film-substrate interface at $z \pe 0$ is coherent.
In the reference state, the film is flat and the elastic energy
density equals $\cEz \pe \Ef \, (a_f-a_s)^2/a_s^2 (1-\nf)$, where
$a_\alpha$, $E_\alpha$ and $\nu_\alpha$ are the lattice parameter,
Young modulus and Poisson ratio of the solid $\alpha$, with $\alpha
\pe f$ for the film, and $s$ for the substrate. The dynamical
equation of the film shape is then, see \cite{Srol89,SpenVoor91},
\begin{equation}
 \label{surfevol}
    \frac{\partial h}{\partial t} = D \sqrt{1+|\nabla
    h|^2} \, \nabla_S^2 \mu \, ,
\end{equation}
with $D$, a constant related to surface diffusion, and $\nabla_S$, the
surface gradient. Both elastic $\Fel$ and surface
$\Fsurf \pe \int d\vecr \gamma (h) \sqrt{1+|\nabla h|^2}$ free energies 
contribute to the surface chemical potential 
$\mu \pe \delta (\Fel + \Fsurf)/\delta h$ which reads
\begin{equation}
\label{defmu}
\mu = \cE [h] + \gamma(h) \kappa(h) +
\gamma'(h)/\sqrt{1+|\nabla h|^2} \, ,
\end{equation}
with $\cE[h]$, the elastic energy density computed at $z \pe h(x,y,t)$, $\gamma$, the isotropic surface energy and $\kappa$, the free surface mean curvature. To account for wetting \cite{ChiuGao93,GoloDavi03,TekaSpen04,PangHuan06}, the surface
energy $\gamma$ is supposed to be a function of the film height extrapolating
from the bulk value $\gamma_f$ when $h \! \rightarrow \! \infty$, to
some upper value when $h \! \rightarrow \! 0$. Here, the wetting 
interactions are described by the characteristic length $\delta$ and the 
strength $\cw \ppg 0$, and the surface energy is written as
$\gamma (h) \pe \gamma_f \left[ 1 + \cw f\left(h/\delta
\right)\right]$ with some function $f$ going to zero at infinity. In the
following we will use $f(\xi) \pe \exp(-\xi)$ when a specific form is needed.
Finally, we set the length unit to be $\lz \pe \cEz / \gamma_f$, the characteristic 
length of the instability with the corresponding time unit $\tz \pe \lz^4/D \gamma_f$.

To compute elastic energies, we use the isotropic continuum framework with
stresses $\sigma_{pq}^\alpha$ proportional to strains $e_{pq}^\alpha$
in the solid $\alpha$,
\begin{equation}
  \sigma_{pq}^{\alpha} =   \frac{E^\alpha}{1+\nu^\alpha}
    \left[ e_{pq}^{\alpha} +\frac{\nu^{\alpha}}{1-2
        \nu^{\alpha}}  e_{nn}^{\alpha}\delta_{pq}
    \right],
\end{equation}
with summation over repeated indices, $n,p,q \pe x,y,z$,
$\delta_{pq}$, the Kronecker symbol, and $e_{pq}^\alpha \pe \fud
(\partial_q u_p^\alpha + \partial_p u_q^\alpha) - \eta^\alpha \,
\delta_{pq} (\delta_{p1}+\delta_{p2}) $ where $\mathbf{u}$ is the
displacement with respect to the reference state commensurate with
the substrate so that $\eta^f \pe a^f/a^s-1$ and $\eta^s \pe 0$.
Since the system is at mechanical equilibrium, it satisfies
$\partial_q \sigma_{pq}^\alpha \pe 0$ with the following boundary
conditions: $\vecu^s \! \rightarrow \mathbf{0}$ when $z \!
\rightarrow \! -\infty$ and is continuous at $z \pe 0$, whereas 
$\sigma_{pz}^\alpha$ is continuous at $z \pe 0$ while
$\sigma_{pq}^f n_q \pe 0$ at the free surface $z \pe h(x,y,t)$ with
the outward normal $\mathbf{n}$. To solve for elasticity, we use the thin 
film approximation \cite{TekaSpen04} assuming the thickness $h$ to be an order 
$\epsilon$ smaller than the characteristic length $\lz$. Hence, in the film, we
consider the rescaled variables $Z \pe z/\epsilon$ 
and get $\mathbf{u}$ considering an expansion up to
$\epsilon^3$, $\mathbf{u} \pe \sum_{n = 0}^3 \epsilon^n
\mathbf{u}^{(n)} (x,y,Z)$. In the plane substrate however,
elasticity is solved as usually using Fourier transforms with
respect to $\vecr \pe \{x,y\}$, $\mathcal{F}[h] \pe (2\pi)^{-2} \int
d \vecr e^{i\vk \cdot \vecr} h(\vecr)$. Eventually, we calculate the
elastic energy $\cE \pe \fud e_{pq}^\alpha \sigma_{pq}^{\alpha}$ 
up to $\epsilon^2$, the first nonlinear term, and
obtain the central equation of this Letter describing the film dynamics,
\begin{multline}
\label{dhadt}
  \frac{\partial h}{\partial \tb} = \Delta
  \left\{ \rule{0mm}{7mm}
    - \left[ 1 + \cw f\left(\frac{h}{\db} \right) \right] \Delta h
        + \frac{\cw}{\db}
        \frac{f'\left(h/\db\right)}{\sqrt{1+|\nabla
        h|^2}} \right. \\
        - \wu  \calH_{ii} (h)
    + \wdd \left(2h \Delta h + \left| \nabla h \right|^2 \right)
    \\\left.
    + \wde \left( 
         2 \calH_{ij}\left[ h \, \theta_{ijkl} \calH_{kl}(h)\right]
         +\calH_{ij}(h) \theta_{ijkl} \calH_{kl}(h)
    \rule{0mm}{5mm}
    \right)
 \rule{0mm}{7mm}  \hspace{-0.2mm} \right\},
\end{multline}
with $i,j,k,l\pe x,y$. In \eqref{dhadt}, we use the notation
$\theta_{ijij} \pe 1$, $\theta_{xxyy} \pe \theta_{yyxx} \pe \nf$,
$\theta_{xyyx} \pe \theta_{yxxy} \pe -\nf$, and $\theta_{ijkl} \pe
0$ otherwise, and define the functionals $\calH_{ij} [h] \pe \tfmu
\{ (k_i k_j/k) \, \cF[h] \}$ with $k \pe |\vk|$. The different elastic constants are
$\wu  \pe 2 \Ef (1-\ns^2) / \Es (1-\nf)$, $\wdd \pe (1+\nf)/(1-\nf)
+ \Ef(1-2\ns)(1+\ns)/\Es(1-\nf)$ and $\wde \pe 2 \Ef^2
(1-\ns^2)^2/\Es^2 (1-\nf)^2 (1+\nf)$, which match $2 (1+\nu^{eq})$
in the case of equal film and substrate elastic properties, $\ns \pe
\nf \pe \nu^{eq}$ and $\Es \pe \Ef \pe E^{eq}$. In the latter case,
Eq.~\eqref{dhadt} coincides with the linear analysis of
\cite{TekaSpen04}. In fact, up to order $h^2$, Eq.~\eqref{dhadt} 
can be derived thanks to the elastic free energy
\begin{multline}
\label{defFel}
 \Fel = \int d\vecr h(\vecr) \left[ -\fud \wu \calH_{ii}(h) - \wdd |\nabla h| ^2
    \right. \\ \left. \rule{0mm}{5mm}
    + \wde \calH_{ij} (h) \theta_{ijkl} \calH_{kl} (h) \right] .
\end{multline}
Finally, in two dimensions (2D), Eq.~\eqref{dhadt} reduces to
\begin{multline}
\label{dhadtdd}
 \frac{\partial h}{\partial \tb} = \frac{\partial^2}{\partial x^2}
\left\{
    - \left[ 1 + \cw f\left(\frac{h}{\db} \right)\right] h_{x x}
    + \frac{\cw}{\db}  \frac{f'\left( h/\db\right)}{\sqrt{1+h_x^2}}
    \right. \\
     - \wu  \calH(h_x) + \wdd \left( 2 h h_{x x} + h_x^2 \right) \\
    \left. + \wde \left( 2 \calH \left\{ \left[ h \calH(h_x) ²\right]_x \right\}
              + \left[ \calH(h_x) \right]^2 \right) \rule{0mm}{6mm} \right\},
\end{multline}
where $x$-indices denote $x$-derivatives and $\calH$ is the Hilbert transform 
acting in Fourier space as $\calH [h_x] \pe \tfmu \left\{ |k| \cF[h] \right\}$. 
In the case of equal film and substrate elastic properties and without wetting,
we retrieve the result of \cite{XianE02} describing a 2D semi-infinite film.

We now investigate the dynamics predicted by Eqs.~\eqref{dhadt} and
\eqref{dhadtdd}. In the linear regime, considering small
perturbations of amplitude  $\exp[\sigma(\vk)t+i \vk \! \cdot \!
\vecr]$ around a flat film of height $\hz$, we find $\sigma(\vk) \pe
- a k^2 + \wu k^3 - b k^4$ with $a \pe 1+\cw f(\hz/\delta)$ and $b
\pe \cw f''(\hz/\delta)/\delta^2$. Hence, when $f''(\xi)$ is
decreasing and positive, there exists some critical height $h_c$
below which $\sigma (\vk) \ppp 0$ everywhere so that the film
is linearly stable thanks to the wetting interactions. However, for
$\hz \ppg h_c$, $\sigma (\vk) \ppg 0$ in a given $k$-interval, and the
film is linearly unstable. For
small wetting length $\delta$ and exponential wetting potential, one
gets $h_c \simeq - \wdd \delta \ln (\wu \delta^2 / 4 \cw)$. When
$\cw \pe 0$, the film is always linearly unstable and the
nonlinear numerical computations exhibit generic finite-time blow-up
solutions \cite{GoloDavi03}.

\begin{figure}[ht] \centering
\includegraphics[width=0.4\textwidth]{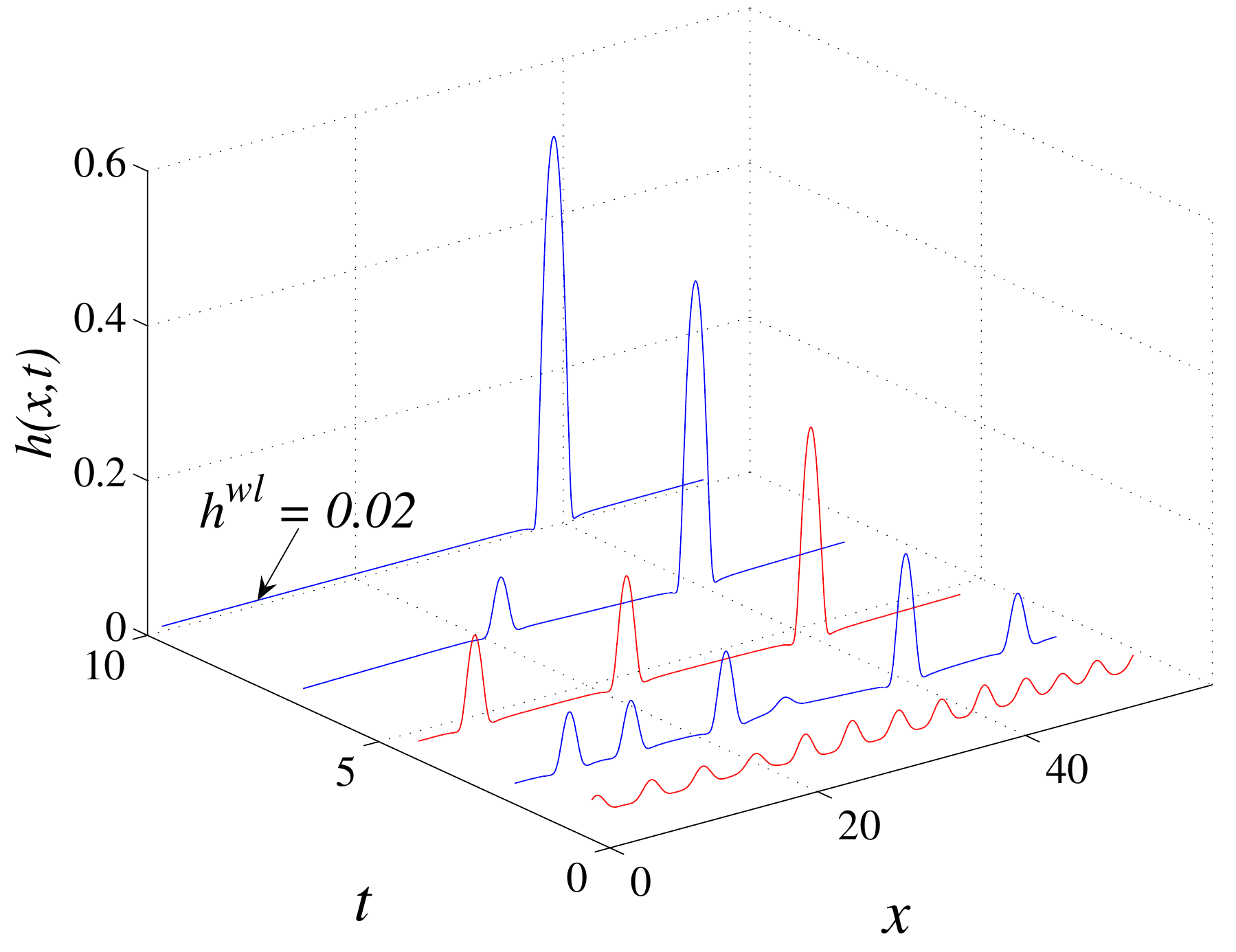}
\caption{Space-time evolution of a 2D film according to
\eqref{dhadtdd} with $\hz \pe 0.1$. Surface diffusion induces a
non-interrupted coarsening until only one island is left surrounded
by a wetting layer with height $\hwl$.} \label{fig1}
\end{figure}

\begin{figure*}[ht] \centering
\includegraphics[width=0.31\textwidth]{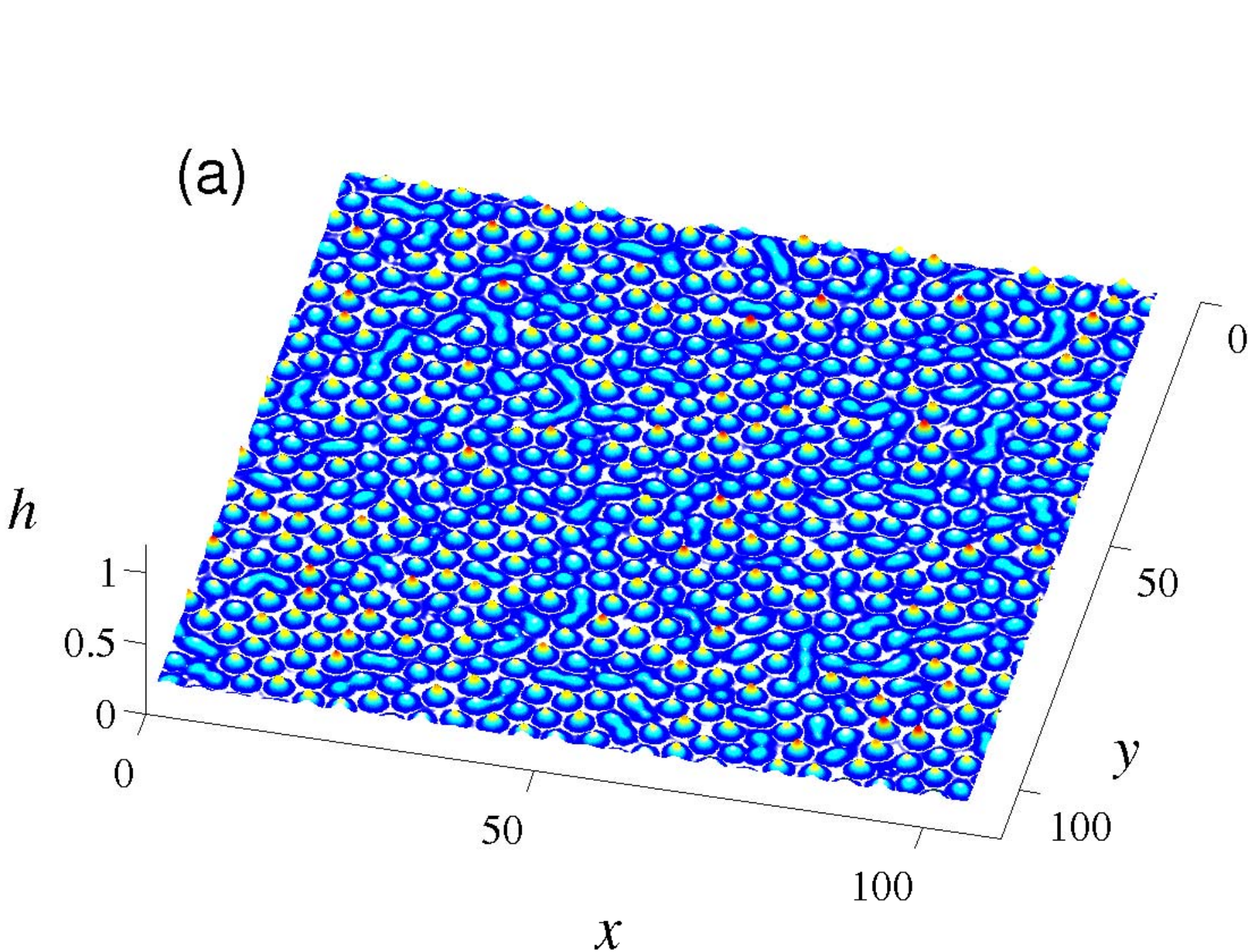}
\includegraphics[width=0.31\textwidth]{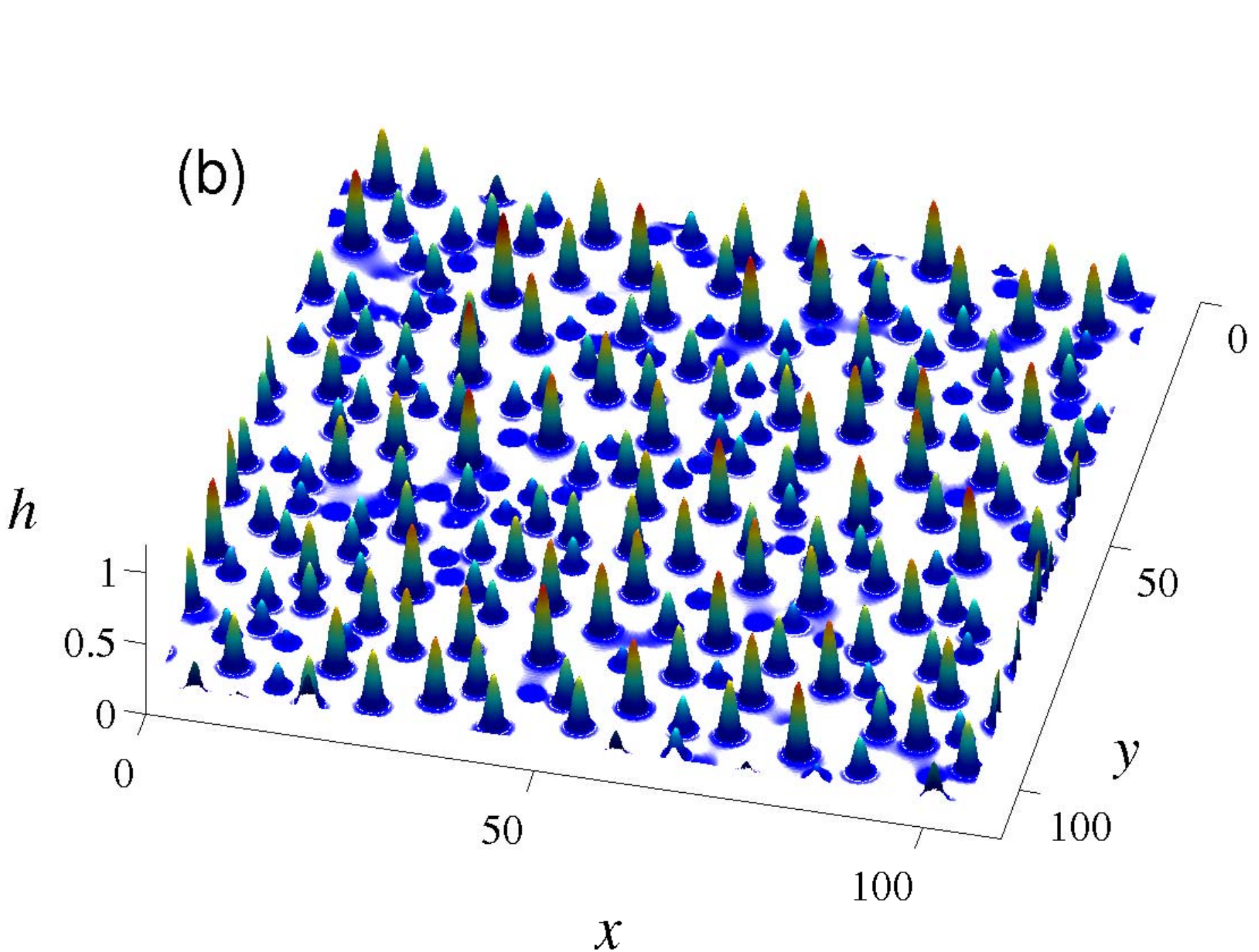}
\includegraphics[width=0.31\textwidth]{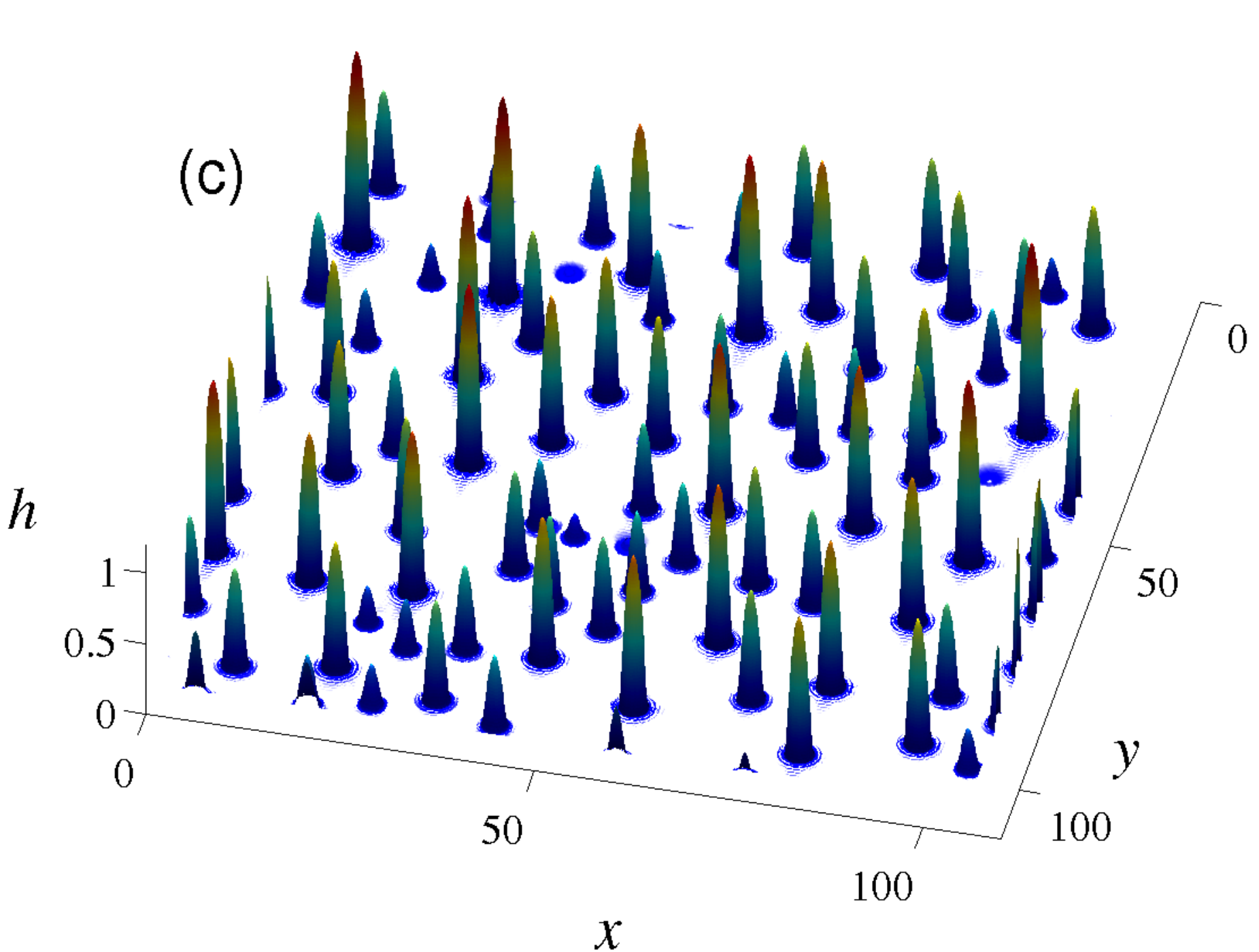}
\caption{Space-time evolution obtained by Eq.~\eqref{dhadt} with $t \pe
0.39$ (a), $0.83$ (b) and $1.39$ (c).} \label{fig2}
\end{figure*}

To analyze the nonlinear evolution in the presence of wetting
effects, we performed numerical simulations using a pseudo-spectral
method in a periodic box of length $L$. To be specific, we selected
parameters depicting a ${\rm{Si}}_{0.8} {\rm{Ge}}_{0.2}$ film on a
Si substrate with $\nf \pe 0.278$, $\wu \pe 2.44$, $\wdd \pe 2.52$
and $\wde \pe 2.34$, leading to $\lz \pe 200\,$nm and
with the value of the diffusion parameter $D$ given in
\cite{SpenVoor91}, $\tz \pe 8\,$hours at $750$°C, see \cite{Flor00}. 
In fact, thanks to space and time rescaling, only $\nf$ and $\wde/\wdd$ 
are relevant for characterizing Eqs.~\eqref{dhadt} and \eqref{dhadtdd}.
The wetting potential is described in an indicative way by $\cw \pe
0.05$ and $\delta \pe 0.005$. The initial condition is a flat film
perturbed by a small noise with a mean initial height $\hz$. As
shown in Figs.~\ref{fig1} and \ref{fig2}, a film with $\hz \ppg h_c$ is
first destabilized by the morphological instability which generates
surface undulations according to the linear growth. The linear 
stage is then quickly replaced by a nonlinear one characterized by the 
emergence of well-defined islands which grow without moving 
and with a decreasing aspect ratio. The islands are
surrounded by a wetting layer smaller than $h_c$ which allows
surface transport and the subsequent islands ripening. Hence,
no singularity appears here in the nonlinear dynamics of the
wetting film. Studying parity in $h$ of the different terms in
\eqref{defFel}, one concludes following \cite{XianE02} that
the last nonlinear nonlocal term drives the
surface towards smooth peaks and deepening and sharpening valleys
which would lead to singularities \cite{XianE02}. Here, the wetting
effects included in $\gamma (h)$ enforce a higher energetic cost for
small $h$ and thus stabilize the thin film. In fact, both nonlocal
nonlinearity and wetting are needed to regularize the dynamics of
the instability which we now characterize by its final state and
time dependence.  Note that a steady island dynamics was also found in 
\cite{PangHuan06} which however did not account consistently for the 
difference in the film and substrate elasticity and lead to 
different long time behavior. 

\begin{figure}[ht] \centering
\includegraphics[width=0.23\textwidth]{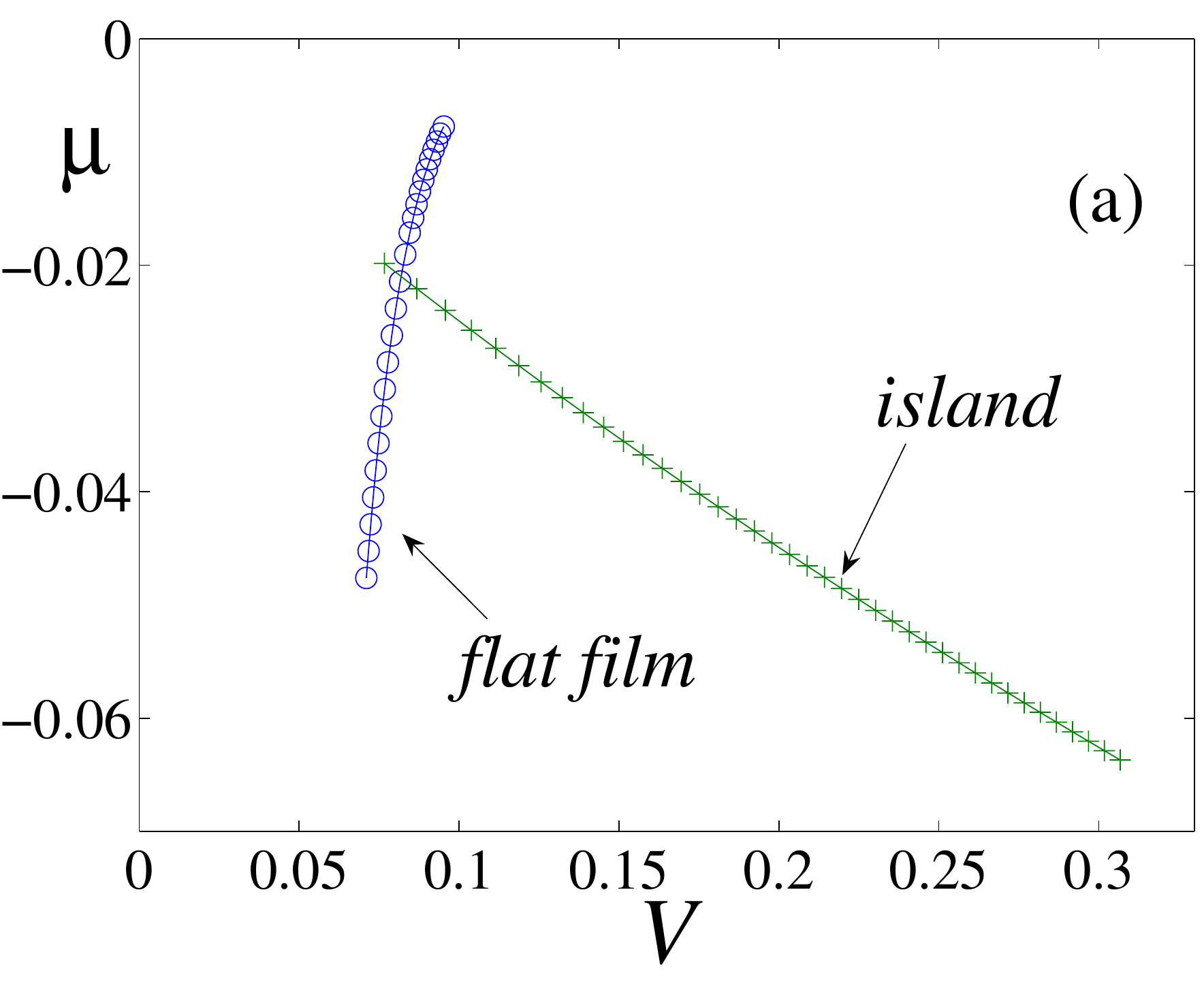}
\includegraphics[width=0.23\textwidth]{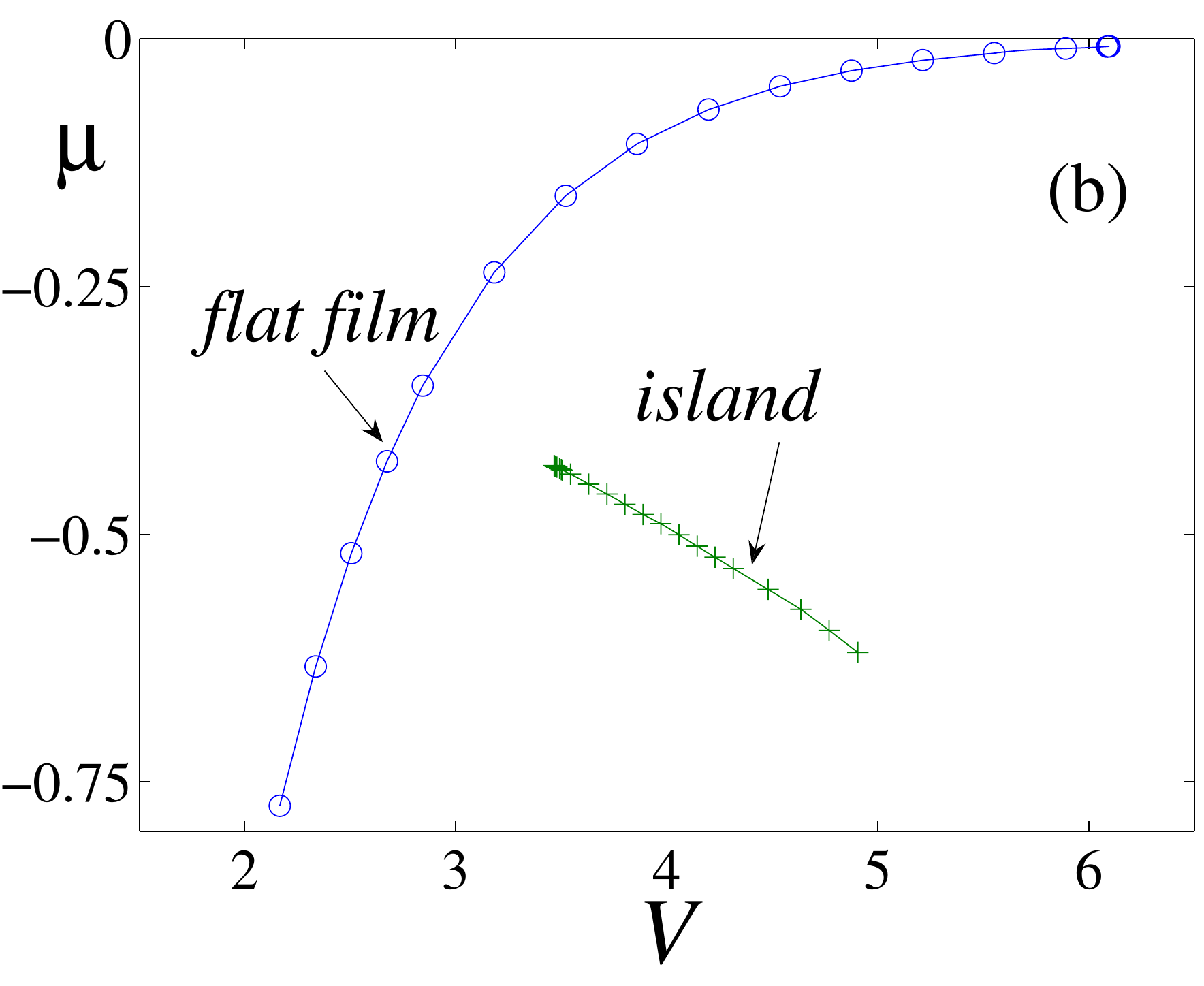}
\caption{Equilibrium surface chemical potential \eqref{defmu} as function 
of the flat film volume $V \pe \Vf$ when $h\ppp h_c$, and of the island volume 
$V$ when $h \ppg h_c$ in 2D (a), 3D (b).} \label{fig4}
\end{figure}

\begin{figure}[ht] \centering
\includegraphics[width=0.23\textwidth]{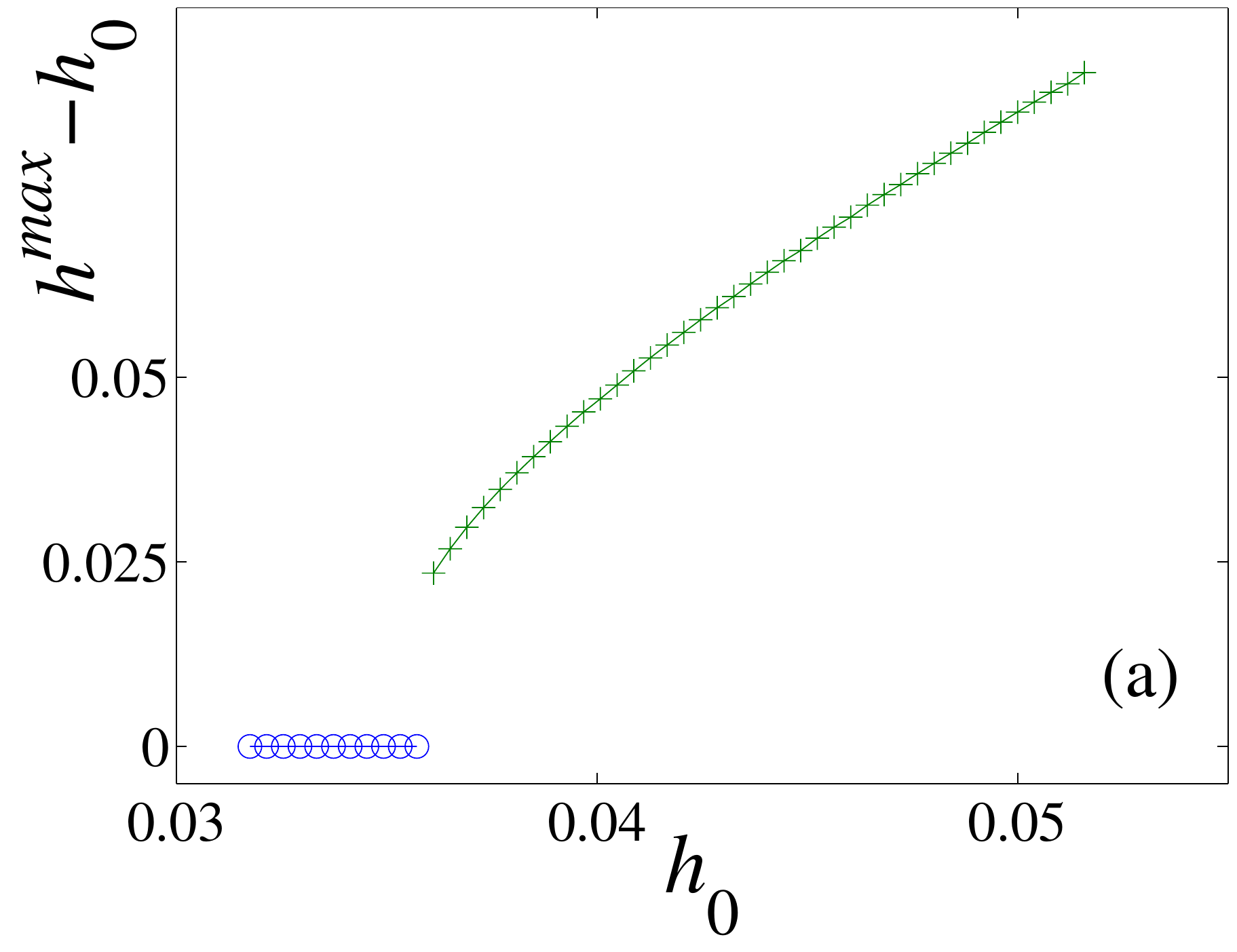}
\includegraphics[width=0.23\textwidth]{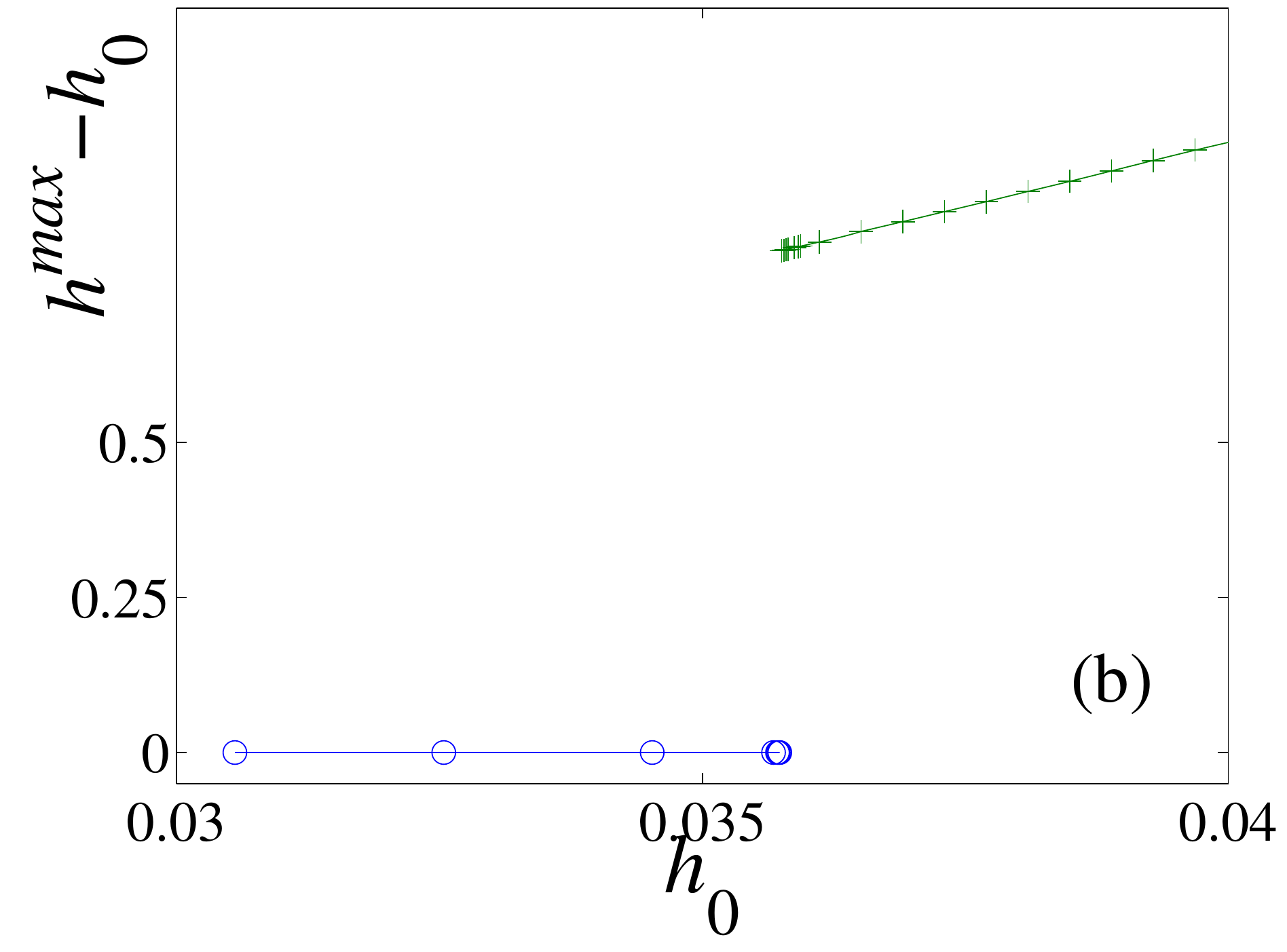}
\caption{Maximal height of an equilibrium island as function of the
initial height in 2D (a), 3D (b).} \label{fig3}
\end{figure}

Within the present model, we observe that the system evolves continuously 
towards an equilibrium state characterized when $\hz \ppg h_c$ by a single 
stable island in equilibrium with a wetting layer of height $\hwl$ in both 
2D and 3D, see e.g. Fig.~\ref{fig1}, whereas when $\hz \ppp h_c$, the final stage is 
a flat film of height $\hwl \pe \hz$. The equilibrium properties ($\hwl$, 
island volume $V$ measured above $\hwl$, etc.) depend only on the homogeneous 
chemical potential \eqref{defmu} and on the sign of $\hz -h_c$ for large enough $L$. 
Computing $\mu$ and $V$ as parametric functions of the film volume $\Vf$, we find that 
when $h \ppp h_c$, $\mu \pe \gamma'(h)$ increases with $\Vf$ until 
$\Vf_c \pe L^2 h_c$, whereas when $h \ppg h_c$, $\mu$ depends only on $V$ and is
monotonously decreasing in both 2D and 3D, see Fig.~\ref{fig4}. Hence, in a regime of 
well-separated islands, bigger ones should always grow by surface diffusion at the 
expense of smaller ones. At equilibrium, we also compute the maximum height $\hmax$ 
as function of the initial height $\hz$, see Fig.~\ref{fig3}. The system undergoes a 
discontinuous bifurcation as the difference $\hmax - \hz$ displays a jump at the
transition height $h_c$ which agrees within a few percents with the
linear estimate $h_c \! \simeq \! 0.036$ corresponding to $7\,$nm.
This first-order like transition also shown in the $\mu(V)$ plot of Fig.~\ref{fig4}
is  at stake in similar instabilities \cite{Nozi93}.

\begin{figure}[ht] \centering
\includegraphics[width=0.23\textwidth]{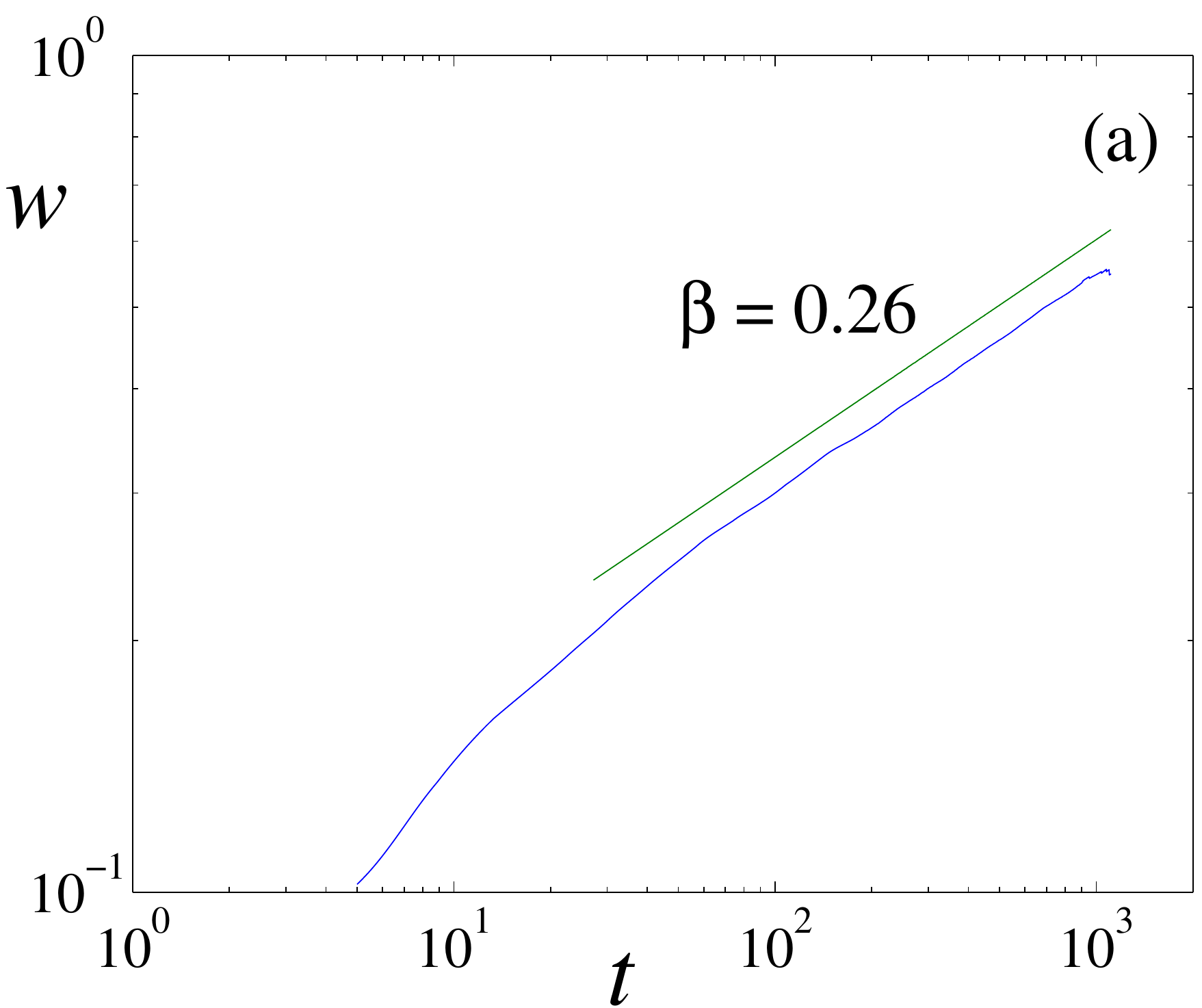}
\includegraphics[width=0.23\textwidth]{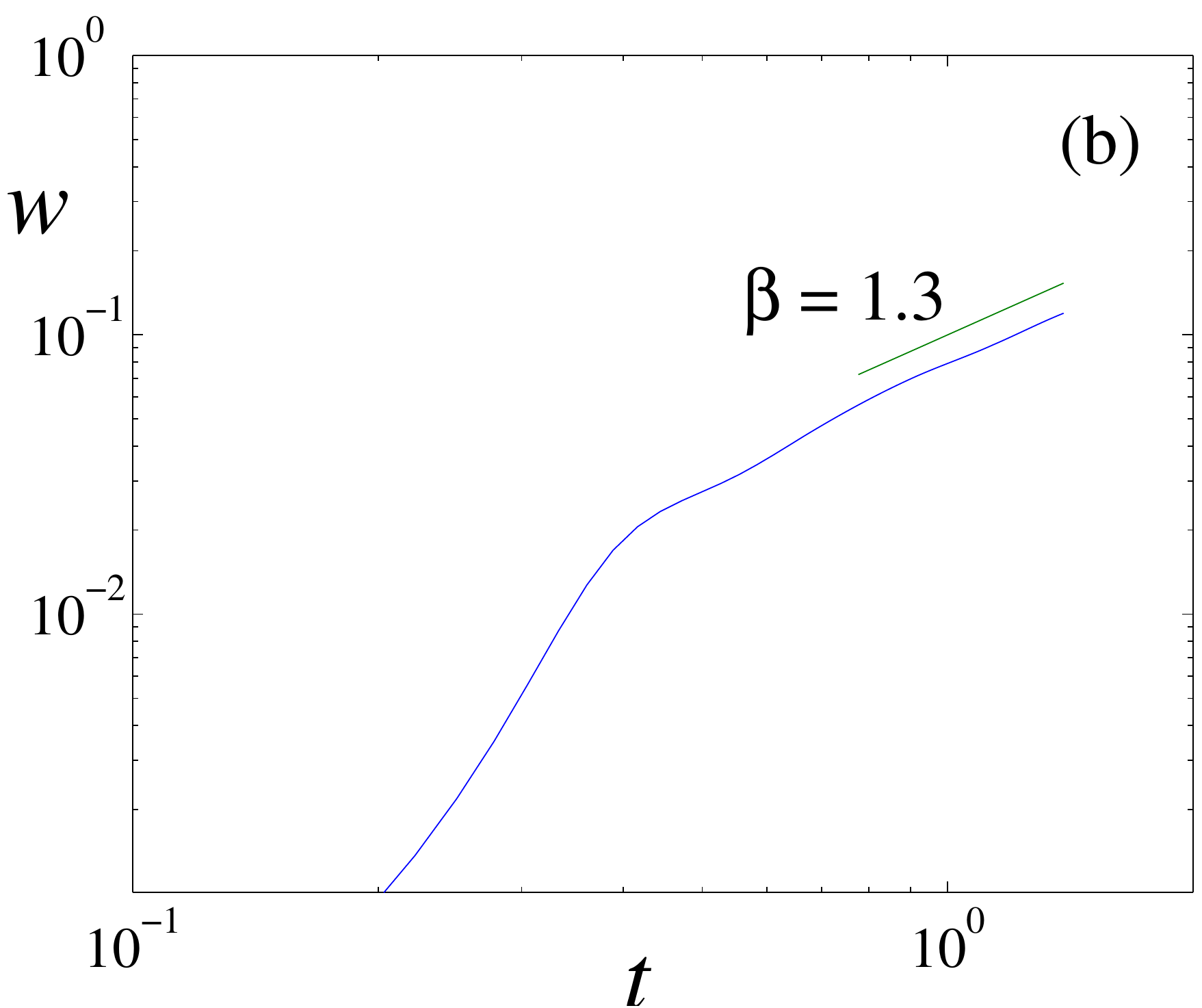}
\caption{Roughness as function of time with $L \pe 6700$ in 2D (a) and $L \pe 104$ in 3D (b).}
\label{fig5}
\end{figure}

\begin{figure}[ht] \centering
\includegraphics[width=0.23\textwidth]{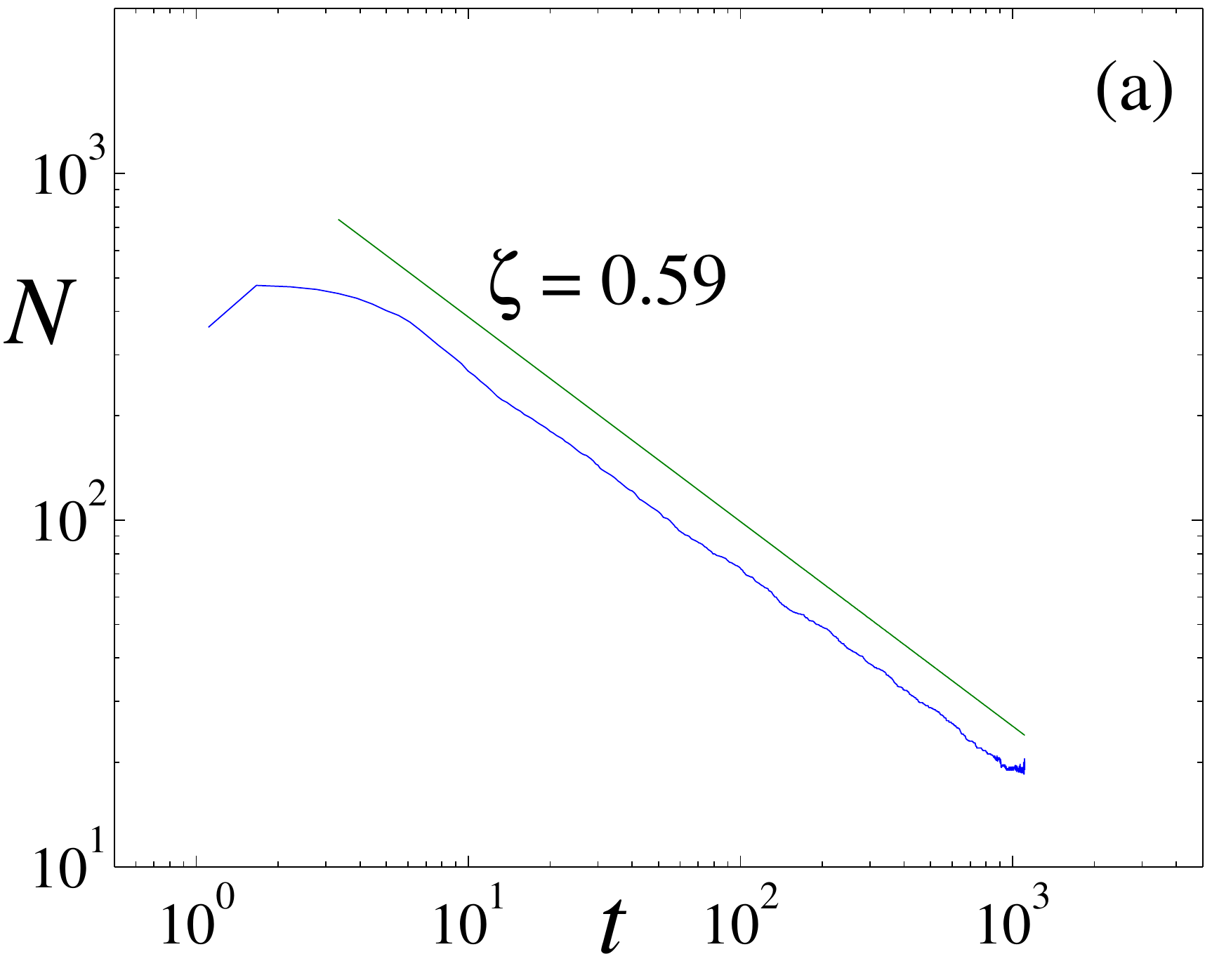}
\includegraphics[width=0.23\textwidth]{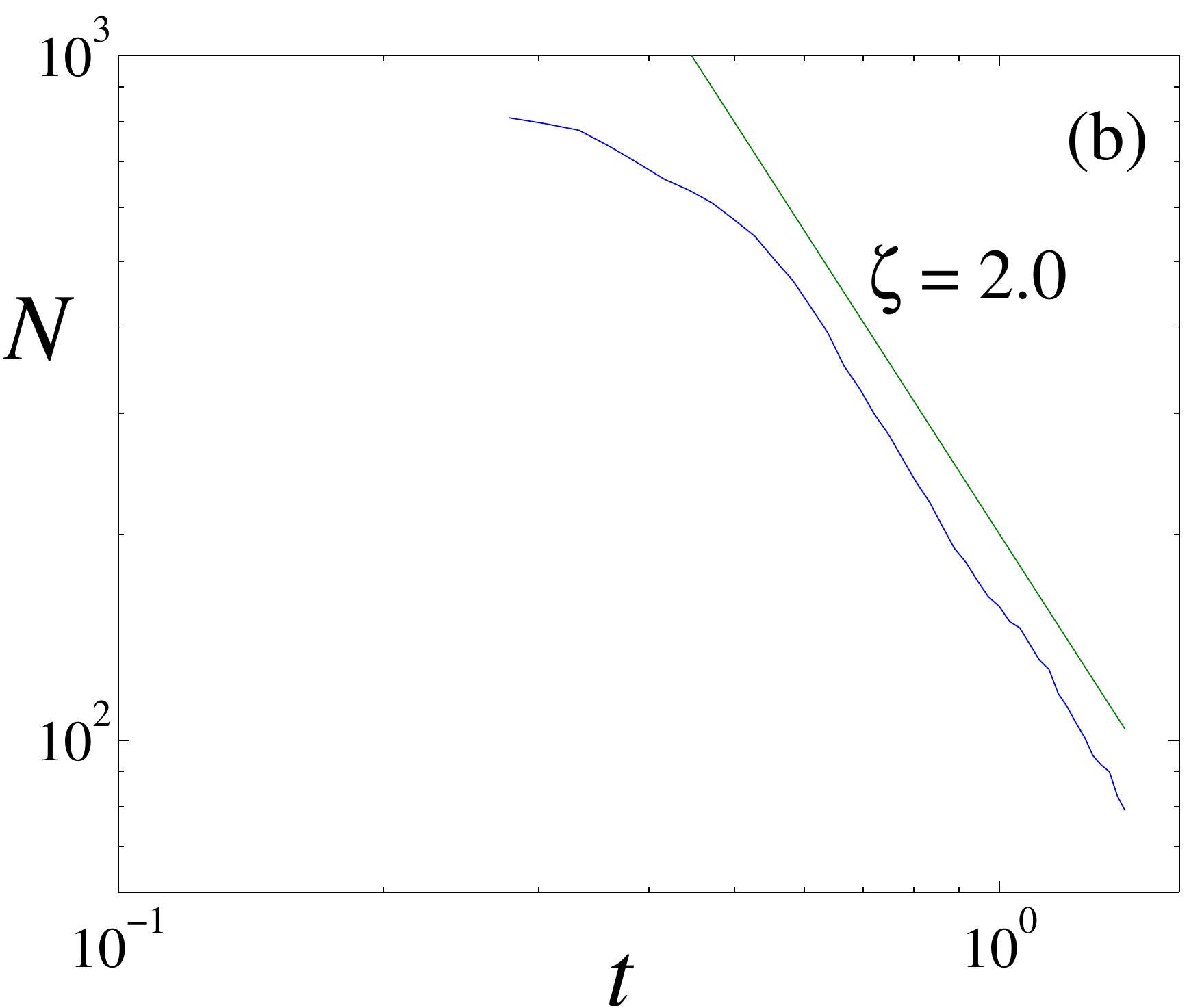}
\caption{Evolution of the number of islands as in Fig.~\ref{fig5}.} \label{fig6}
\end{figure}

Finally, to describe the dynamics of the island growth, we compute the surface
roughness $w(t) \pe [\langle h^2 \rangle -\langle h \rangle^2 ]^{1/2}$
and number of islands $N(t)$. Both 2D and 3D simulations
reveal a non-interrupted coarsening with power-law
behavior $w(t) \psim t^\beta$ and $N(t) \psim 1/t^\zeta$, see 
Figs.~\ref{fig5} and \ref{fig6}. For 2D systems, we find  
$\beta \pe 0.26$ and $\zeta \pe 0.59$ over nearly three decades. Similarly,
over the last time-decade of the 3D simulations, we find $\beta \pe
1.3$ and $\zeta \pe 2.0$ which are noticeably departing from the 2D
values, illustrating the difference between diffusion process over
a one or two dimensional surface. If one sought a self-similar
solution of Eqs.~\eqref{dhadt} or \eqref{dhadtdd} as 
$h \! \sim \! t^\beta \hat{H}(\vecr/t^{1/z})$ for large $t$, one would 
get $\beta \pe 1/3$, thence requiring a more elaborate theory accounting 
for the wetting layer and nonlocal nonlinearity.

In summary, we derived nonlinear and nonlocal equations describing the stress
driven morphological instability of a thin film on a deformable
substrate with a priori different elastic constants and which account for
wetting interactions. When both nonlocal nonlinearity and wetting
are present, numerical simulations reveal a steady
evolution towards an equilibrium state contrarily to the crack 
solutions predicted for the bulk morphological instability.
When the film initial height is higher than some critical value given by 
the wetting interactions, the final stage consists of a single island with a 
chemical potential monotonously decreasing with its volume. Consistently, the
system undergoes a non-interrupted coarsening in both two and three
dimensions characterized by a power-law decrease of the island
number with time which strongly depends on the system
dimensionality. Further experiments on the number of islands 
of annealing films in the prepyramid regime of the Stranski-Krastanov mode 
\cite{Flor00} would be of great interest. New effects such as
anisotropy and faceting will be included in future work.

\begin{acknowledgments}
The authors thank I. Berbezier, J. Villain, P. Müller, A.
Sa\'ul, L. Raymond and P. Meunier for fruitful discussions and assistance.
Support from the \textsc{ANR} via the grant PNANO-M\'EMOIRE, is 
acknowledged.
\end{acknowledgments}


\end{document}